\documentclass[epj]{svjour}
\usepackage{graphics}
\begin{document}
\title{Chiral SU(3) dynamics, $\bar{K}N$ interactions\\ and the quest for antikaon-nuclear clusters}
\author{Wolfram Weise}
%
%
\institute{Physik-Department, Technische Universit\"at M\"unchen, D-85747 Garching, Germany}
\date{Received: date / Revised version: date}
%
\abstract{
This presentation reviews recent developments in the understanding of low-energy kaon-nucleon interactions as they relate to the possible existence of antikaon-nuclear quasibound states. A state-of-the-art  discussion of low-energy $\bar{K}N$ interactions is given, with special emphasis on the subthreshold region relevant to the proposed kaon-nuclear systems. }
\maketitle
\section{Introduction and outlook}
\label{intro}
The low-energy interactions of kaons with nuclear systems are governed by the spontaneous and explicit breaking of chiral $SU(3)\times SU(3)$ symmetry in QCD. Spontaneous chiral symmetry breaking assigns the role of Goldstone bosons to the octet of light pseudoscalar mesons. Explicit symmetry breaking by the small but non-vanishing masses of the light quarks shifts the masses of these mesons to their observed positions. The strange quark mass, $m_s\sim 0.1$ GeV, can still (with caution) be considered small compared to the characteristic scale of spontaneous chiral symmetry breaking, $\Lambda_\chi = 4\pi f \sim 1$ GeV, expressed in terms of the pseudoscalar decay constant, $f\simeq 0.09$ GeV. 

Given this symmetry breaking pattern of low-energy QCD, the leading (Tomozawa-Weinberg) interactions of kaons and antikaons with nucleons are determined unambiguously. In particular, the driving $\bar{K}N$ interaction in the isospin $I=0$ channel is strongly attractive around threshold, $\omega \simeq m_K$. Early discussions of kaon condensation in dense matter \cite{KapNel} were based on this observation which, ever since, has given rise to speculations about  the possible existence of antikaon-nuclear bound states.

The recent revival of this theme was prompted by Akaishi and Yamazaki
\cite{AY02} who used a simple potential model to calculate and predict bound states of few-body systems such as $K^-pp$, $K^-ppn$ and $K^-pnn$. It was argued that if the binding is sufficiently strong to fall below the $\bar{K}N\rightarrow \pi\Sigma$ threshold, such states could be narrow. An experiment performed at KEK with stopped $K^-$ on $^4$He \cite{Su04} seemed indeed to indicate such deeply bound narrow structures with binding energies $B(K^-ppn) \simeq 169$ MeV, $B(K^-pnn) \simeq 194$ MeV and widths $\Gamma < 20$ MeV.  However,  the subsequent repetition of this experiment with better statistics \cite{Iwa06} did not confirm the previously published results. The FINUDA measurements with stopped $K^-$ on $^{6,7}$Li and $^{12}$C targets \cite{Ag05} suggested an interpretation in terms of quasibound $K^-pp$ clusters with binding energy $B(K^-pp) = (115 \pm 9)$ MeV and width $\Gamma = (67\pm 16)$ MeV. However, this interpretation was criticized in Ref.\cite{OT06} with the argument that the observed spectrum may be explained by final state interactions of the produced $\Lambda p$
pairs. Another line of experimental studies focuses on the invariant mass spectroscopy of $\Lambda p$
pairs produced in heavy-ion collisions at GSI and analyzed with the FOPI detector \cite{GSI06}. 

At present, the issue of deeply bound $K^-$-nuclear states is still unresolved. On the theoretical side, reliable calculations need to be performed which require a detailed treatment of three basic ingredients: 
\begin{itemize}
\item{\bf Realistic $\bar{K}N$ interactions}
\end{itemize}
Chiral SU(3) dynamics with coupled channels involving the $\Lambda$(1405) resonance, plus $p$-wave interactions dominated by the $\Sigma$(1385), provide a useful framework for extrapolations into the relevant range below $\bar{K}N$ threshold. These extrapolations are presently still subject to uncertainties which will be progressively reduced by more accurate threshold data. 
\begin{itemize}
\item{\bf Realistic $NN$ interactions}
\end{itemize}
The repulsive short-range nucleon-nucleon interaction works against the strong compression of $\bar{K}$-nuclear systems proposed in Ref.\cite{AY02} and must be handled appropriately. 
\begin{itemize}
\item{\bf Realistic $\bar{K}NN \rightarrow \Lambda N, \Sigma N$ absorption}
\end{itemize}
Apart from the $\bar{K}N\rightarrow\pi\Sigma, \pi\Lambda$ widths, $\bar{K}$ absorption processes on two nucleons are the primary source of imaginary parts which tend to prohibit narrow bound states.  So far, not much is known empirically about these absorptive widths. This is key information for which only (exclusive) experiments can provide reliable answers in the future.

The present article is intended to give a state-of-affairs discussion of theoretical estimates. Our own work in progress in this area is pursued together with Akinobu Dot$\acute{\rm e}$ \cite{DW06} and Rainer H\"artle \cite{HW06}.

\section{Chiral SU(3) dynamics and \\low-energy $\bar{K}N$ interactions}
\label{chiral}
Chiral perturbation theory (ChPT) as a systematic expansion in small momenta and quark masses
is limited to low-energy processes with light quarks. It is an interesting issue to what extent 
the generalisation of ChPT including strangeness can be made to work. The $\bar{K} N$ channel is of particular interest in this context, as a testing ground for chiral SU(3) symmetry in QCD and for the role of explicit chiral symmetry breaking by the strange quark mass. However, any perturbative approach breaks down in the vicinity of resonances. 
In the $K^- p$ channel, for example, the existence of the $\Lambda(1405)$ resonance 
just below the $K^- p$ threshold renders SU(3) ChPT
inapplicable. At this point the combination with non-perturbative coupled-channels
techniques has proven useful, by generating the $\Lambda(1405)$ dynamically
as an $I=0$ $\bar{K} N$ quasibound state and as a resonance in the $\pi \Sigma$
channel \cite{KSW95}. Coupled-channels methods combined with chiral $SU(3)$ dynamics have subsequently been applied to a variety of meson-baryon scattering processes with quite some success \cite{KWW97}. A recent update is given in \cite{BNW05}. 

The starting point is the chiral $SU(3) \times SU(3)$ meson-baryon effective Lagrangian. Its leading order terms include the octet of pseudoscalar Goldstone bosons ($\pi, K, \bar{K}, \\ \eta$) and their interactions. Symmetry breaking mass terms introduce the light quark masses $m_u, m_d$ and the mass of the strange quark, $m_s$. The pseudoscalar mesons interact with the baryon octet ($p, n, \Lambda, \Sigma, \Xi$) through vector and axial vector combinations of their
fields. At this stage the parameters of the theory, apart from the pseudoscalar meson decay constant $f \simeq 90$ MeV, are the $SU(3)$ baryon axial vector coupling constants
$D \simeq 0.80$ and $F \simeq 0.47$ which add up to $D + F = g_A = 1.27$. At next-to-leading order, seven additional constants enter in $s$-wave channels, three of which are constrained by mass splittings in the baryon octet and the remaining four need to be fixed by comparison with low-energy scattering data.

\subsection{Coupled channels} 
Meson-baryon scattering amplitudes based on the $SU(3)$ effective Lagrangian involve coupled channels for each set of quantum numbers. For example, The $K^- p$ system in the isospin $I=0$ sector couples strongly to the $\pi\Sigma$ channel. 
Consider the $T$ matrix ${\bf T}_{ij}(p, p')$ connecting meson-baryon channels $i$ and $j$ with four-momenta $p, p'$ in the center-of-mass frame:
\begin{eqnarray}
{\bf T}_{ij}(p, p') &=&
 {\bf K}_{ij}(p, p') \nonumber\\
 &+& \sum_n\int{d^4q\over (2\pi)^4} {\bf K}_{in}(p, q)
 \,{\bf G}_n (q)\,{\bf T}_{nj}(q, p')\,\, 
\label{T}
\end{eqnarray}
where ${\bf G}$ is the Green function describing the intermediate meson-baryon loop which is iterated to all orders in the integral equation\footnote{Dimensional regularisation with subtraction constants is used in practise.}. The driving terms ${\bf K}$ in each channel are constructed from the chiral $SU(3)$ meson-baryon effective Lagrangian in next-to-leading order. In the kaon-nucleon channels, for example, the leading terms have the form\footnote{The convention for the T matrix used here differs from the (dimensionless) one in Ref.\cite{BNW05} by a factor $(2M_N)^{-1}$.}
\begin{equation}
{\bf K}_{K^\pm p} = 2\,{\bf K}_{K^\pm n}  = \mp {\omega\over f^2} + ...\,\, ,
\label{K}
\end{equation}
at zero three-momentum, where the invariant c.m. energy is $\sqrt{s} = \omega + M_N$ and $f$ is the pseudoscalar meson decay constant. Scattering amplitudes are related to the $T$ matrix (\ref{T}) by ${\bf F} = (M_N/4\pi\sqrt{s})\,{\bf T}$.
Note that ${\bf K} > 0$ means attraction, as seen for example in the $K^- p \rightarrow K^- p$ channel. Similarly, the coupling from $K^- p$ to $\pi\Sigma$ provides attraction, as well as the diagonal matrix elements in the $\pi\Sigma$ channels. Close to the $\bar{K}N$ threshold, we have a leading-order piece   $F(K^- p \rightarrow K^- p) \simeq (1+m_K/M_N)^{-1}\,m_K/4\pi f^2$. This is the analogue of the Tomozawa-Weinberg term (proportional to $m_\pi/4\pi f^2$ in pion-nucleon scattering at threshold), but now with an attractive strength considerably enhanced by the much larger kaon mass $m_K$.
 
When combining chiral effective field theory with the coupled-channels scheme, the "rigorous" chiral counting in powers of small momenta is abandoned in favor of iterating a subclass of loop diagrams to ${\it all}$ orders. However, the substantial gain in physics compensates for the sacrifice in the chiral book-keeping.  Important non-perturbative effects are now included in the re-summation, and necessary conditions of unitarity are fulfilled. 

\subsection{S-wave interactions} 
\begin{figure}
\resizebox{0.45\textwidth}{!}{
\includegraphics{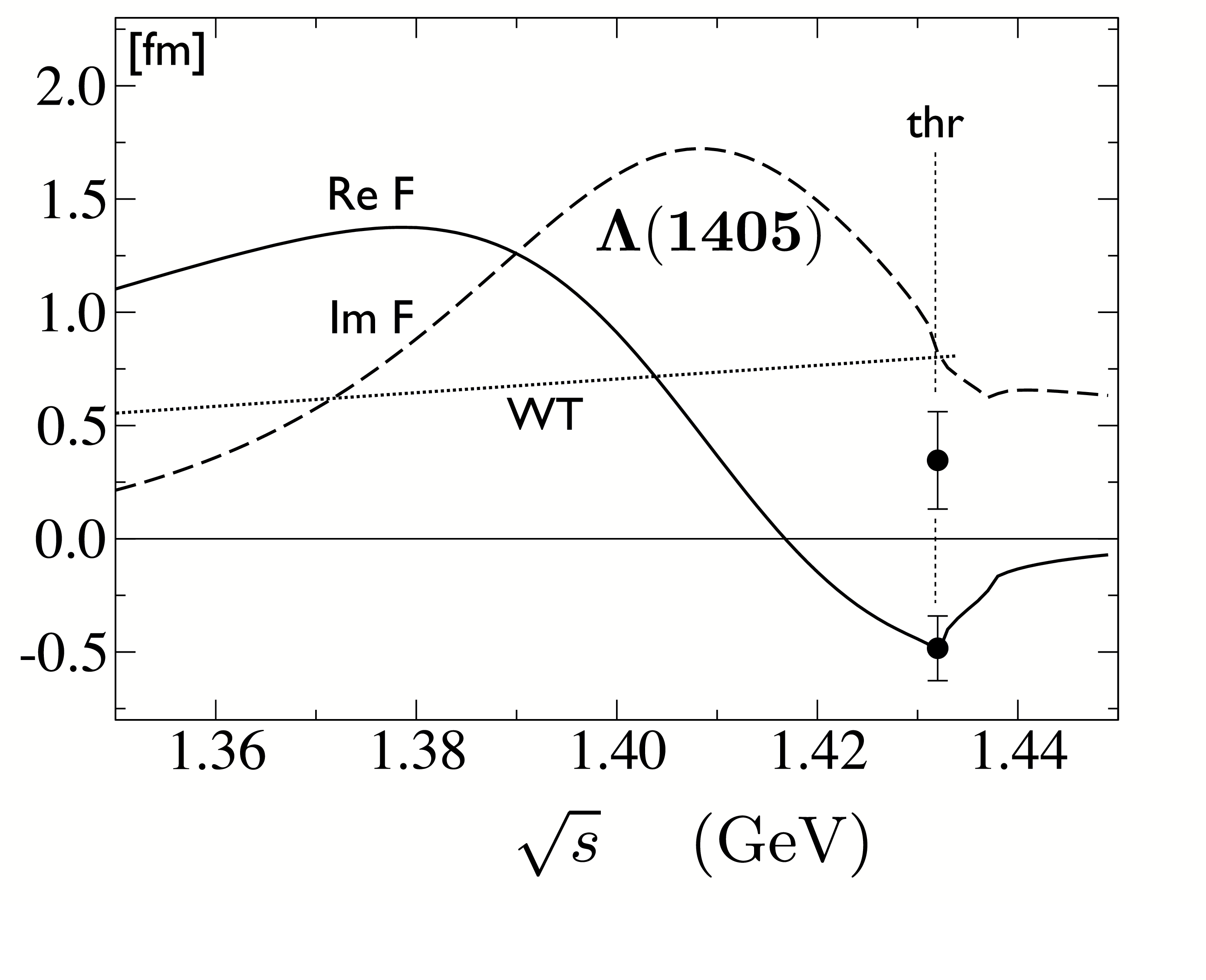}
}
\caption{Real and imaginary parts of the $K^- p$ forward scattering amplitude calculated in the chiral SU(3) coupled channels approach \cite{BNW05}, as functions of  the invariant $\bar{K}N$ center-of-mass energy$\sqrt{s}$. Real and imaginary parts of the scattering length deduced from the DEAR kaonic hydrogen measurements \cite{Beer04} are also shown. The dotted line indicates the leading-order (Tomozawa-Weinberg) amplitude for comparison.}
\label{fig:1}      
\end{figure}

The $K^- p$ threshold data base has recently been improved by new accurate results for the strong interaction shift and width of kaonic hydrogen \cite{Beer04}. These data, together with existing information on $K^- p$ scattering, the $\pi\Sigma$ mass 
spectrum and measured $K^- p$ threshold decay ratios, set tight constraints on the theory and have consequently revived the interest in this field. Fig.\ref{fig:1} shows results of a calculation which combines driving terms from the next-to-leading order chiral $SU(3)$ meson-baryon Lagrangian with coupled-channel equations \cite{BNW05}. As in previous calculations of such kind, the $\Lambda(1405)$ is generated dynamically as an $I = 0$ $\bar{K}N$ quasibound state and a resonance in the $\pi\Sigma$ channel. 

The improved accuracy of the recent kaonic hydrogen data from the DEAR experiment indicate a possible inconsistency with older $K^-p$ scattering data (see Ref. \cite{BNW05}). Note that the real part of the $K^-p$ amplitude, when extrapolated into the subthreshold region below the $\Lambda(1405)$, is expected to be large and positive (attractive).  The imaginary part of this amplitude drops at energies below the $\Lambda(1405)$. The dominant $I = 0$ decay into $\pi\Sigma$ is turned off below its threshold at $\sqrt{s} \simeq 1.33$ GeV. The s-wave $K^-n$ subthreshold amplitude, not shown here, is also attractive but less than half as strong as the $K^-p$ amplitude and non-resonant \cite{BNW05}.  

The off-shell $s$-wave $K^- p$ amplitude resulting from the coupled-channel calculation \cite{BNW05} can be conveniently para-metrised as follows:
\begin{eqnarray}
F_{K^- p}^{s-wave} &=& \left({M_N\over \sqrt{s}}\right)\,{\omega + a\,m_K^2 + b\,\omega^2\over 4\pi f^2} \cdot \nonumber\\
&\cdot&\left(1+{\sqrt{s}\,\gamma_0\over M_0^2 - s -i\sqrt{s}\,\Gamma_0(s)}\right)~~,
\label{swave} 
\end{eqnarray}
with $f \simeq 0.11$ GeV, $a = -b \simeq 1$ GeV$^-1$, $\gamma_0 \simeq 0.21$ GeV and the $\Lambda(1405)$ mass and width ($M_0, \Gamma_0$). This form is useful for practical purposes and reflects the behavior of the leading and next-to-leading order terms as well as the non-perturbative part involving the dynamically produced resonance.  

\subsection{An equivalent pseudopotential}

In applications to nuclear systems it is often useful to translate the leading $\bar{K}N$ $s$-wave interaction into an equivalent potential in the laboratory frame (where the nucleon is approximately at rest). The leading order piece (the Tomozawa-Weinberg term) can be viewed as resulting from vector meson exchange \cite{KW}. Starting from the non-linear sigma model in SU(3), introduce gauge couplings of the vector meson octet to the pseudoscalar octet and fix a universal vector coupling constant $g\simeq 6$ such the $\rho \rightarrow \pi^+\pi^-$ width is reproduced. Then construct vector meson couplings to the SU(3) octet baryons through their conserved vector currents.  The corresponding piece of the reduced $\bar{K}N$ interaction Lagrangian which generates the t-channel vector meson exchange $\bar{K}N$ amplitude at tree level, with vector meson mass $m_V$, is
\begin{eqnarray}
\delta{\cal L}(\bar{K}N)&=& {ig^2\over 4}\left(K^-\partial_\mu K^+ - K^+\partial_\mu K^-\right)\cdot \nonumber \\
& &\left[\partial^2 + m_V^2\right]^{-1} \,\bar{\Psi}_N\,\gamma^\mu(\tau_3 + 3)\,\Psi_N~~,
\end{eqnarray}
where $K^{\pm}$ are the charged kaon fields and $\Psi_N = (p,n)^T$ is the isodoublet nucleon field. The isovector ($\tau_3$) piece comes from $\rho$ exchange and the isoscalar part (with its typical factor of 3) comes from $\omega$ exchange, while $\phi$ exchange does not contribute as long as there are no strange quark components in the nucleon. Taking the long-wavelength limit $|\vec{q}|\rightarrow 0$, one arrives at the scattering operator 
\begin{eqnarray}
\delta\hat{T} = (g^2/2m_V^2)\,\omega\, \Psi_N^\dagger\,(\tau_3 + 3)\,\Psi_N~~.\nonumber
\end{eqnarray}
Using the KSFR relation $m_V = \sqrt{2} f g$, the Tomozawa-Weinberg amplitudes (\ref{K}) follow immediately.

These considerations suggest a characteristic range $r\sim m_V^{-1}$ of the $\bar{K}N$ interaction even for $pointlike$ kaon and nucleon. When the actual size of the nucleon is taken into account,
the minimal range of the s-wave $\bar{K}N$ interaction is determined by the form factors related to the vector currents of the nucleon, for which the electromagnetic form factors of the proton are a good measure. A conservative estimate of this range is therefore given by the r.m.s. proton charge radius, $r\sim 0.9$ fm. This is presumably a lower limit since the intrinsic size of the kaon adds to the overall size of the interaction range.

The static pseudopotentials which approximate the $\bar{K}N$ interaction in r-space follow through the operator identity $\hat{V} = -\delta\hat{T}/2\omega$. The result is:
\begin{equation}
V_{K^-p}(\vec{r}) = - {g_p(\vec{r})\over 2f^2}~,~~V_{K^-n}(\vec{r}) = - {g_n(\vec{r})\over 4f^2}~~,
\label{pot}
\end{equation}
with distributions $g_{p,n}(\vec{r})$ normalised to unity. In the limit $m_V\rightarrow\infty$ and for pointlike nucleons,  $g_{p,n}(\vec{r}) \rightarrow \delta^3(\vec{r})$. Applications in few-body calculations commonly use Gaussian forms for $g(\vec{r})$, with range parameters left free and usually chosen smaller than the ``minimal" required input, namely the r.m.s. radius related to the nucleon's vector current.  

With $f\simeq 0.1$ GeV, the potential $V_{K^-p}(\vec{r})$ is not sufficiently strong to produce a quasibound state. In the work of Ref.\cite{AY02}, the coupling strength was roughly doubled in order to generate the $\Lambda(1405)$ at the right place, and needed to be even further amplified to deal with the (at that time still observed) candidates for deeply bound $K^-NNN$ states. This procedure can be misleading. In fact any approach which tries to generate $K^-$nuclear states from a purely static, energy-independent $\bar{K}N$ potential misses important physics, for the following reasons.

The $\bar{K}N\leftrightarrow\pi\Sigma$ coupling is well known to be strong. The measured threshold branching ratios for $K^-p$ into $\pi^\pm\Sigma^\mp$ represent about $2/3$ of all $K^-p$ inelastic channels. Moreover, the large fraction of double charge exchange, $\Gamma(K^-p\rightarrow\pi^+\Sigma^-)/\Gamma(K^-p\rightarrow\pi^-\Sigma^+) \simeq 2.4$, demonstrates the importance of coupled-channel dynamics beyond leading orders. An effective potential, projected into the diagonal $K^-p$ channel, that accounts for these mechanisms will be complex and strongly energy dependent.  

Consider for simplified demonstration a schematic two-channel model involving the coupled $I=0$ states $|1\rangle = |K^-p\rangle$ and  $|2\rangle = |\pi\Sigma\rangle$. We ignore the (relatively unimportant)
$K^-p\leftrightarrow\bar{K}^0 n$ charge exchange channel and let $ |\pi\Sigma\rangle$ stand for all combinations of charges $\pi^\pm\Sigma^\mp$ and  $\pi^0\Sigma^0$. The channel coupling matrix element $V_{12} = \langle K^-p|V|\pi\Sigma\rangle= V_{21}^\dagger$ is of the same order as the diagonal elements $V_{11} = \langle K^-p|V|K^-p\rangle$ and $V_{22} = \langle\pi\Sigma|V|\pi\Sigma\rangle$. Let  $h_{1,2}$ include masses and kinetic energies in the respective channels and let the wave function of the coupled system be written $|\psi\rangle = c_1\, |K^-p\rangle +   c_2\, |\pi\Sigma\rangle$:
\begin{eqnarray}
(h_1 + V_{11} - E)\, c_1&=& -V_{12}\, c_2~~,\nonumber \\
(h_2 + V_{22} - E)\, c_2 &=& -V_{21}\, c_1~~.
\label{cc}
\end{eqnarray}
The primary mechanism for generating the $\Lambda(1405)$ is resonance formation in the $\pi\Sigma$ channel. Assume therefore that the uncoupled equation of motion for $|\pi\Sigma\rangle$ produces a pole at $E = m_0 - (i/2)\Gamma_0$. The $\Lambda(1405)$ with its physical mass $M_0$ is then supposed to emerge as a $K^-p$ (quasi-)bound state embedded in the $\pi\Sigma$ continuum once the channel coupling is turned on. Eliminating $c_2$ from Eq.(\ref{cc}), the remaining equation for $c_1$ projected into the $K^-p$ channel obviously involves the complex and energy dependent effective potential
\begin{equation}
V_{eff}(E) = V_{11} - {|V_{12}|^2\over m_0-{i\over 2}\Gamma_0 -E}~~.
\end{equation} 
Such a non-local $K^-p$ interaction $V_{eff}(E)$ is to be used in variational calculations which do not treat the $\pi\Sigma$ channels explicitly. The form of this potential resembles the one of the coupled-channels amplitude (\ref{swave}). For an attractive $V_{11} < 0$ and at energies below the $\pi\Sigma$ resonance, $E < m_0$, the attractive strength of $V_{eff}$ can easily become twice that of $V_{11}$ itself, as seen also in the $K^-p$ amplitude, Fig.\ref{fig:1}.    

\subsection{P-wave interactions}

P-waves play a minor role in $\bar{K}N$ interactions close to threshold. However, as pointed out in Ref.\cite{GW05}, they are of potential importance for tightly bound $\bar{K}$-nuclear systems in which the antikaon can have large three-momentum. A useful parametrisation of these amplitudes\footnote{This is an update of the form given long ago in Ref.\cite{BWT78}.} involves dominantly the $\Sigma(1385)$ resonance accompanied by a small background term:
\begin{equation}
F_{K^- p}^{p-wave} = {1\over 2}F_{K^- n}^{p-wave}=  {M_N\over \sqrt{s}}\,C(s)\,\vec{q}\cdot \vec{q}' ~~,
\label{p1} 
\end{equation}
\begin{equation}
C(s) = {\sqrt{s}\,\gamma_1\over M_1^2 - s -i\sqrt{s}\,\Gamma_1(s)} + d~~,
\label{p2} 
\end{equation}
with $\gamma_1 \simeq 0.42/m_K^2$, $d \simeq 0.06$ fm$^3$, $M_1 = 1.385$ GeV and  (energy dependent) width $\Gamma_1\simeq 40$ MeV at resonance. Note that these $p$-wave amplitudes
represent attractive interactions below the $\Sigma(1385)$. Here the isospin $I=1$ dominates so that, unlike the $s$-wave case, the $K^-$neutron interaction is now twice as strong as that for $K^-$proton. 

\section{Antikaon-nuclear bound states}

\subsection{The $K^-pp$ system}

The present theoretical investigations of possible $K^-pp$ quasibound states use two complementary approaches: the variational AMD (Antisymmetrized Molecular Dynamics) method \cite{AY02,DW06} and three-body coupled-channel Faddeev calculations \cite{SGM06}.

The most recent studies using the AMD framework \cite{DW06} start from a realistic $NN$ interaction (Argonne v18) together with energy dependent $s$- and $p$-wave $\bar{K}N$ interactions based on Eqs.(\ref{swave},\ref{p1},\ref{p2}). These calculations are still preliminary and subject to improvements in treating the short-range repulsive $NN$ correlations in the variational $K^-pp$ wave function. The first results indicate that the total $K^-pp$ binding energy does not exceed about 50 MeV and the short range $NN$ repulsion prevents strong compression of the system, unlike the earlier suggestions of Ref.\cite{AY02}. First estimates indicate that the total width of this state is larger than 100 MeV, about  20$\%$ of which is expected to come from $K^-NN\rightarrow YN$ absorption.

The Faddeev calculations \cite{SGM06} are performed using separable $NN$, $\bar{K}N$ and $YN$
interactions and include $\bar{K}N \leftrightarrow\pi\Sigma$ channel coupling. The input parameters are constrained by properties of the $\Lambda(1405)$ and by low-energy scattering data where available.
Depending on details of the parameter sets, the calculated pole positions of the three-body T-matrix in the complex plane give the following range for binding energy and width of $K^-pp$:
\begin{eqnarray}
B(K^-pp) &\sim& (55 - 70)\,{\mbox{MeV}}\,,\nonumber \\
\Gamma(K^-pp\rightarrow\pi\Sigma N) &\sim& (95 - 110)\, {\mbox{MeV}}~. 
\end{eqnarray} 
The absorptive width $\Gamma(K^-pp\rightarrow YN)$, not included in these computations, would add to increase the total width well beyond 100 MeV. While these first exploratory variational and Faddeev calculations are roughly consistent amongst themselves, they are (so far) not compatible with the interpretation of the FINUDA data \cite{Ag05} as signals for the formation of deeply bound $K^-pp$ clusters with binding energy as large as $B(K^-pp) \sim 115$ MeV and a width around 70 MeV.

\subsection{Antikaons in nuclear matter}

Kaonic nuclei with a $K^-$ bound to heavier nuclear cores are likewise of interest, although their experimental detection would certainly be difficult. As a generic starting point of this discussion,
consider $K^-$ modes in nuclear matter. The kaon spectrum in matter with proton and neutron densities $\rho_{p,n}$ is determined by
\begin{equation}
\omega^2 - \vec{q}\,^2 - m_K^2 - \Pi_K(\omega,\vec{q};\rho_{p,n}) = 0,
\label{Keqn}
\end{equation} 
with the $K^-$ self-energy $\Pi_K$ (or equivalently, the $K^-$ nuclear potential $U_K$) in the nuclear medium: 
\begin{equation}
\Pi_{K^-} = 2\omega \,U_{K^-} = -\left[T_{K^-p}\,\rho_p + T_{K^-n}\,\rho_n\right] + \, ...
\end{equation} 
where $T_{K^-p,n}$ are the $K^-p$ and $K^-n$ (forward scattering) T-matrices. The additional terms, not shown explicitly, include corrections from Fermi motion, Pauli blocking, two-nucleon correlations etc.
An effective kaon mass in the medium can be introduced by solving Eq.(\ref{Keqn}) at zero momentum: $m_K^*(\rho) = \omega(\vec{q} = 0, \rho)$. 

\begin{figure}
\resizebox{0.45\textwidth}{!}{
\includegraphics{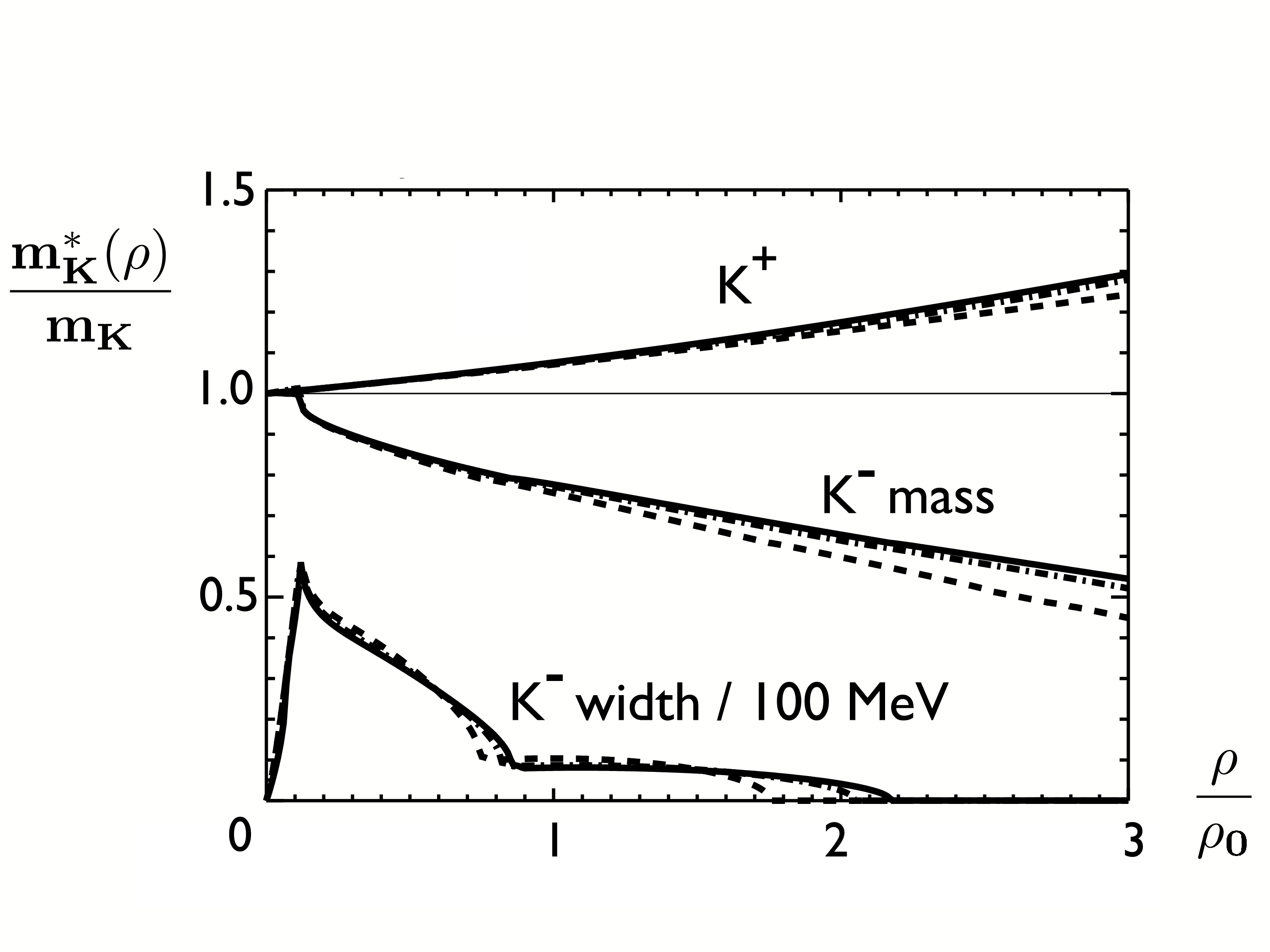}
}
\caption{In-medium mass $m_K^*(\rho)$ and width of a $K^-$ in symmetric nuclear matter as a function of baryon density $\rho$ in units of nuclear matter density $\rho_0 = 0.17$ fm$^{-3}$. The calculations \cite{WKW96} were performed using in-medium chiral SU(3) dynamics combined with coupled channels and including effects of Pauli blocking, Fermi motion and two-nucleon correlations. Also shown is the in-medium $K^+$ effective mass calculated in the same approach.}
\label{fig:2}      
\end{figure}

Calculations of the spectrum of kaonic modes as a function of density have already a long history. For example, in Refs. \cite{WKW96} it was pointed out that, as a consequence of the underlying attractive $\bar{K}N$ forces, the $K^-$ mass at the density of normal nuclear matter ($\rho_0 \simeq 0.17$ fm$^{-3}$) effectively drops to about three quarters of its vacuum value. At this density the $K^-$ in-medium decay width is expected to be strongly reduced because the $K^- N$ energy ``at rest" in matter has already fallen below the $\pi\Sigma$ threshold. These calculations do, however, not include the $\bar{K}NN\rightarrow YN$ absorptive width $\Gamma_{abs}$ which grows with $\rho^2$, the square of the baryonic density. A rough estimate \cite{HW06} gives $\Gamma_{abs} \sim 30$ MeV at $\rho = \rho_0$ which adds to the width shown in Fig.\ref{fig:2}. 

\subsection{Kaonic nuclei}

Mares et al. \cite{MFG06} have recently studied the possiblity of $\bar{K}$-nuclear bound states using a relativistic mean field model in which the $\bar{K}$ couples to scalar and vector fields mediating the nuclear interactions. Estimates of the absorptive width are also made. Kaon-nuclear binding energies are found in the range $B_K \sim 100 - 200$ MeV accompanied by widths with a lower limit of about 50 MeV.

An alternative, exploratory calculation \cite{HW06}, using a realistic subthreshold $\bar{K}N$ interaction as described in Section 2, can be based on the Klein-Gordon equation with a complex, energy dependent
$K^-$ self-energy $\Pi_K(\omega, \vec{r})$. Bound states are determined as eigenstates of

\begin{equation}
[\omega^2 + \vec{\nabla}^2 - m_K^2 - {\mbox{Re}}\Pi_K(\omega,\vec{r})]\,\phi_K(\vec{r}) = 0~~,
\label{KGeqn}
\end{equation}
where the Coulomb interaction is introduced by the gauge invariant relacement $\omega \rightarrow \omega + V_c(\vec{r})$. The width of the bound state is calculated according to
\begin{equation}
\Gamma = -{1\over\omega}\int d^3r \,\phi_K^*(\vec{r})\, {\mbox{Im}}\Pi_K\,\phi_K(\vec{r})
\end{equation} 
The kaon self-energy includes $s$- and $p$-wave interactions to leading order in density:
\begin{eqnarray}
\Pi_K(\omega, \vec{r}) &=& \Pi_s(\omega, \vec{r}) + \Pi_p(\omega, \vec{r}) + \Delta\Pi_K~~,\\
\Pi_s(\omega, \vec{r}) &=& - 4\pi\left(1 + {\omega\over M_N}\right)\cdot\nonumber \\ 
&\cdot&\left[ F_{K^-p}(\omega)\,\rho_p(\vec r\,)+  F_{K^-n}(\omega)\,\rho_n(\vec r\,)\right] ~~,\nonumber\\
\Pi_p(\omega, \vec{r}) &=& 4\pi\left(1 + {\omega\over M_N}\right)^{-1}\cdot\nonumber \\ 
&\cdot&\vec{\nabla}\left[ C_{K^-p}(\omega)\,\rho_p(\vec r\,)+  C_{K^-n}(\omega)\,\rho_n(\vec r\,)\right]\vec{\nabla} ~~.\nonumber
\end{eqnarray} 
The $s$- and $p$-wave amplitudes, $F(\omega)$ and $C(\omega)$, are given by Eqs.(\ref{swave}, \ref{p1}, \ref{p2}) and $\Delta\Pi_K$ stands for a series of higher-order corrections (Pauli and short-range corelations, two-nucleon absorption etc.). Pauli and short-range correlations including charge exchange channels are dealt with in a way analogous to the method described in Ref.\cite{WRW96} for nuclear matter, but now transcribed using local density distributions. 

The proton and neutron densities $\rho_p(\vec{r}) = \rho_0 (Z/A) w_p(\vec{r})$ and $\rho_n(\vec{r}) = \rho_0 (N/A) w_n(\vec{r})$ are parametrized in term of Wood-Saxon type distributions $w(\vec{r})$ normalized to unity. The central density $\rho_0$ is varied in order to examine the effects of a possible compression of the core nuclei. 

The influence of two-nucleon absorption processes on the bound state widths is estimated introducing an absorptive piece 

\begin{equation}
\Delta\Pi_{abs} = -4\pi i\,B_0\left(1+{\omega\over 2M_N}\right)\left(\rho_p^2 + 2\rho_p\rho_n + {1\over 3}\rho_n^2\right)~~.
\label{abs}
\end{equation}
The reduced $\rho_n^2$ term approximately takes into account the fact that $K^-$ absorption on a neutron pair can only lead to a single $\Sigma^-n$ final state whereas absorption on $pp$ and $pn$ pairs generates $\Sigma N$ and $\Lambda N$ with a greater variety of charge combinations. The value of $B_0$ is subject to large uncertainties. For orientation we use $B_0 \sim 1$ fm$^4$ guided by constraints from the widths of kaonic atom states \cite{MFG06}.

\begin{figure}
\resizebox{0.45\textwidth}{!}{
\includegraphics{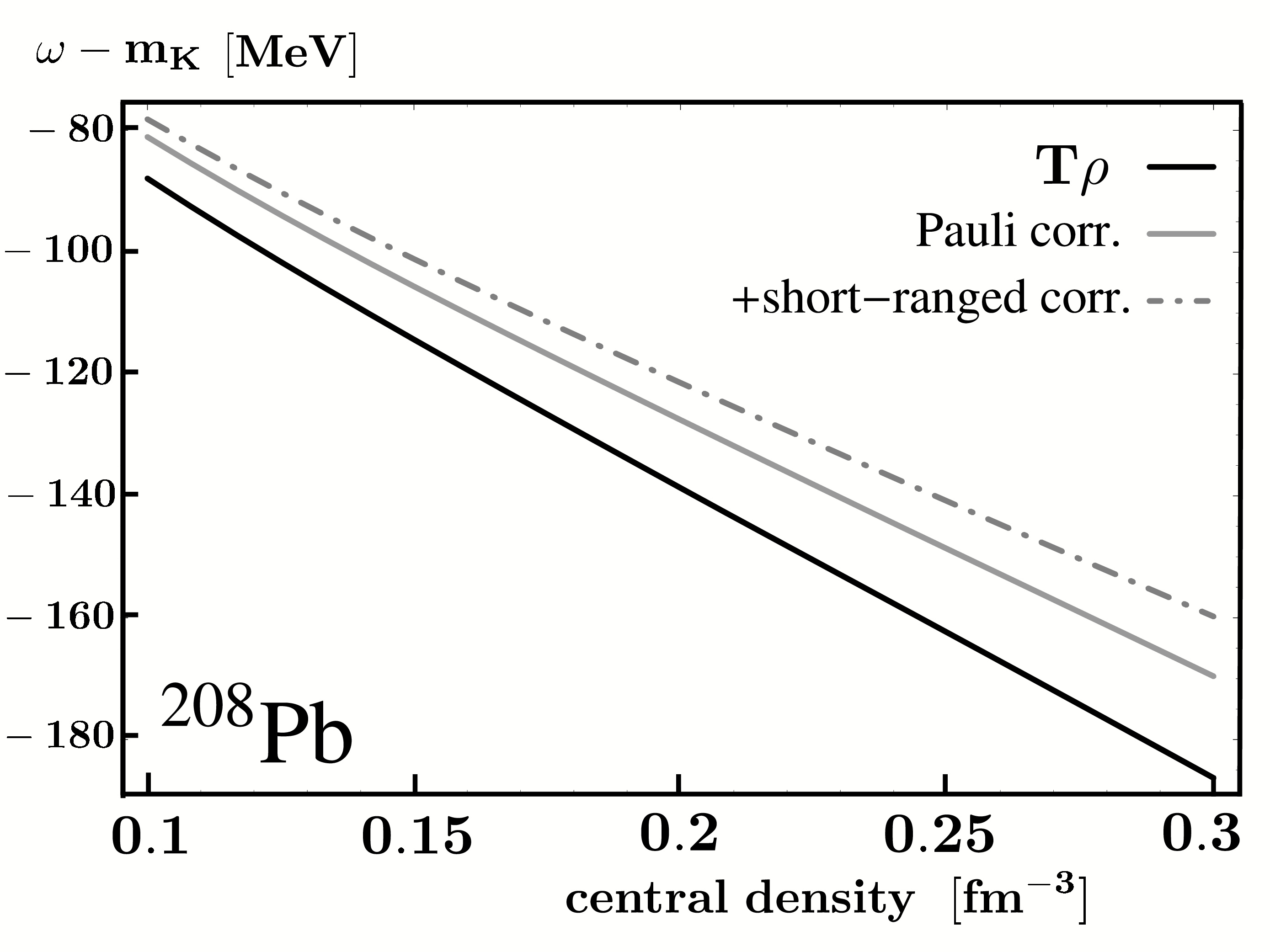}
}
\end{figure}
\begin{figure}
\resizebox{0.45\textwidth}{!}{
\includegraphics{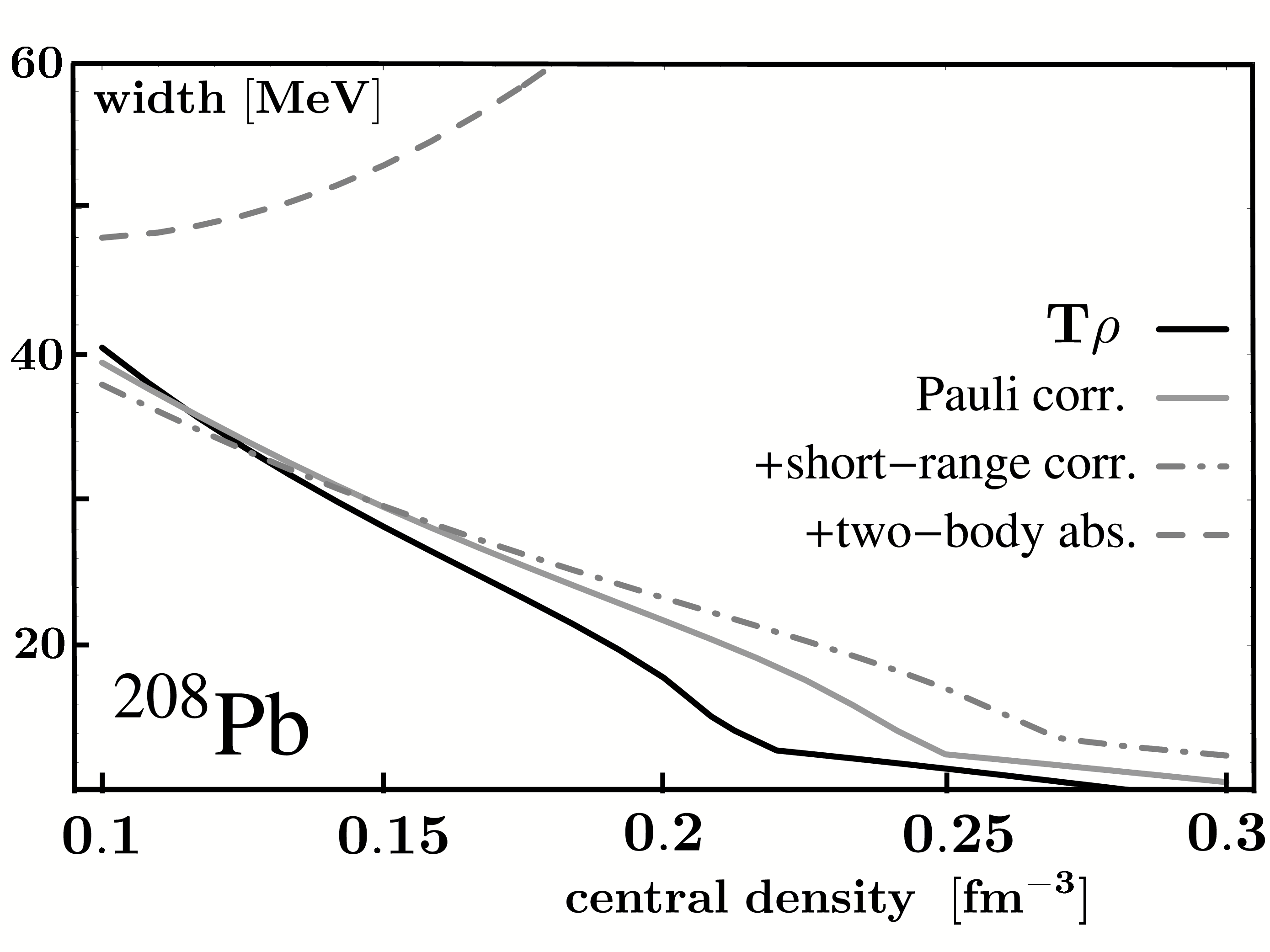}
}
\caption{Binding energy $-B = \omega - m_K$ (upper panel) and width (lower panel) of $K^-$ bound in $^{208}Pb$. The curves represent leading order $s$- and $p$-wave interactions ($T\rho$) and show the effects of Pauli and short-range NN correlations as indicated. The dashed curve in the lower panel gives an impression of the increased width when $K^-NN\rightarrow YN$ absorption is incorporated. Adapted from Ref.\cite{HW06}.}
\label{fig:3}      
\end{figure}

Representative results \cite{HW06} for a kaonic nucleus with a $K^-$ bound in $^{208}Pb$ are shown in Fig.\ref{fig:3}. The leading $s$-wave interaction produces strong antikaon binding. The $p$-wave interaction tends to increase the binding only marginally for $^{208}Pb$ but has a more pronounced effect in lighter nuclei. Pauli and short-range repulsive correlations tend to reduce the binding, as expected. The width is strongly enhanced with increasing density when $K^-NN \rightarrow YN$ absorption is included. This enhancement of the width is more pronounced than in \cite{MFG06} where the absorption term was parametrized as linear (rather than quadratic) in the density.

\subsection{Concluding remarks}

The issue of deeply bound antikaon-nuclear systems (``kaonic nuclei") is a very interesting one but so far unsettled. Early model calculations of kaonic few-nucleon systems did not yet use realistic $\bar{K}N$ and $NN$ interactions. More recent computations with improved interactions come to the (tentative) conclusion that $K^-pp$ as a prototype of an antikaon-nuclear cluster is not as deeply bound as anticipated and presumably has a very short lifetime corresponding to a width of more than 100 MeV. The previously published narrow $K^-NNN$ signals have now disappeared in a measurement with much improved statistics. Deeply bound $K^-$ states in heavier nuclei may exist, but with large widths. A more detailed understanding of these widths  and their underlying mechanisms calls for systematic, exclusive measurements of the final states resulting from $K^-$ induced processes, especially in light nuclei.  

\vspace{0.3cm}
Stimulating discussions with Avraham Gal and Paul Kienle are gratefully acknowledged.
This work is supported in part by BMBF and GSI.

\end{document}